\documentclass[aps,reprint,groupedaddress,nofootinbib, prl]{revtex4-1}
\usepackage{dcolumn}
\usepackage{graphicx}
\usepackage{color}

\begin{document}
	
\title{Ultracold molecules for measuring the electron's electric dipole moment}
	
\author{J. Lim}
\author{J. R. Almond}
\author{M. A. Trigatzis}
\author{J. A. Devlin}
\author{N. J. Fitch}
\author{B. E. Sauer}
\author{M. R. Tarbutt}
\email{m.tarbutt@imperial.ac.uk}
\author{E. A. Hinds}
\affiliation{Centre for Cold Matter, Blackett Laboratory, Imperial College London, Prince Consort Road, London SW7 2AZ, UK}

\begin{abstract}
	We demonstrate one-dimensional sub-Doppler laser cooling of a beam of YbF molecules to 100~$\mu$K. This is a key step towards a measurement of the electron's electric dipole moment using ultracold molecules. We compare the effectiveness of magnetically-assisted and polarization-gradient sub-Doppler cooling mechanisms.  We model the experiment and find good agreement with our data. 
\end{abstract}
	
\maketitle
	
Molecules are increasingly important for testing fundamental physics. They are used to probe parity violation in nuclei~\cite{Cahn2014} and chiral molecules~\cite{Daussy1999, Tokunaga2013}, search for changing fundamental constants~\cite{Shelkovnikov2008, Hudson2006, Bethlem2008, Bethlem2009, Truppe2013}, test quantum electrodynamics~\cite{Salumbides2011}, and measure the electric dipole moments of electrons~\cite{Hudson2011, Baron2014, Cairncross2017} and protons~\cite{Hunter2012}. Measurements of the electron's electric dipole moment (eEDM) using molecules now provide tight constraints on the parameters of theories that extend the Standard Model~\cite{Nakai2017}.  For an atom or molecule with unpaired electrons, the interaction of the eEDM with an applied electric field induces a linear Stark shift, which through relativistic interactions~\cite{Schiff1963} can greatly exceed that of the bare electron~\cite{Sandars1965,Flambaum1976}. This enhancement is proportional to the degree of polarization, so is generically much larger for polar molecules than for atoms~\cite{Sandars1975, Hinds1997}. Strong polarisation also suppresses important systematic errors arising from motional magnetic fields and geometric phases~\cite{Hudson2002}. In some molecules, the polarization can be reversed by state selection, which is helpful for avoiding systematic errors~\cite{Meyer2006, Vutha2010, Eckel2013}. The first molecular determination of the eEDM used a beam of YbF molecules to obtain an upper limit of $|d_{e}| < 10.5 \times 10^{-28}~e$~cm~\cite{Hudson2011}. A second beam experiment using ThO molecules improved on this, yielding $|d_{e}| < 9.4 \times 10^{-29}~e$~cm~\cite{Baron2014,Baron2017}, and an experiment using trapped HfF$^+$ ions recently gave a similar limit~\cite{Cairncross2017}. 

The linewidth of such an eEDM measurement, or any spectroscopic measurement, cannot exceed the inverse of the coherence time, which for a molecular beam is limited by the thermal expansion of the cloud to $\tau_{\rm max} \simeq \sigma_{\rm max} \sqrt{m/(k_{B}T)}$. Here, $m$ is the molecular mass, $T$ is the translational temperature, and $\sigma_{\rm max}$ is the useable size of the molecular cloud, limited by the detection area or other geometric constraints. So far, eEDM measurements using molecular beams produced at $T \approx 4$~K by supersonic expansion or buffer gas cooling have been limited to $\tau_{\rm max} \approx 1$~ms~\cite{Hudson2011,Baron2014}. A significant improvement requires a much lower temperature, suggesting the need for laser cooling. Recently, laser cooling has been applied to a few molecular species. First, a beam of SrF was cooled transversely and both Doppler and Sisyphus cooling forces were demonstrated~\cite{Shuman2010}. This beam was then slowed by radiation pressure~\cite{Barry2012} and captured and cooled in a magneto-optical trap~\cite{Barry2014, McCarron2015, Norrgard2016, Steinecker2016}. A beam of YO molecules has been cooled, compressed and slowed~\cite{Hummon2013, Yeo2015}, and CaF molecules have been slowed~\cite{Zhelyazkova2014, Truppe2017, Hemmerling2016}, magneto-optically trapped~\cite{Truppe2017b, Williams2017, Anderegg2017} and cooled below the Doppler limit~\cite{Truppe2017b}. Recently, polyatomic SrOH molecules were cooled using Sisyphus forces~\cite{Kozyryev2017}. So far however, laser cooling has not been applied to the heavy polar molecules needed for EDM measurements, though several proposals have been made~\cite{Hunter2012,Tarbutt2013,Isaev2016, Kozyryev2017b}.  Here, we advance towards an eEDM experiment using ultracold molecules by cooling a beam of YbF below $100\,\mu$K, so that a coherence time exceeding $150$~ms is feasible in a beam, a fountain~\cite{Tarbutt2013, Cheng2016} or a trap~\cite{Tarbutt2009}. We have observed Doppler and sub-Doppler cooling, and focus here on the sub-Doppler results. 

	\begin{figure*}[bt]
		\centering
		\includegraphics[width=1\textwidth]{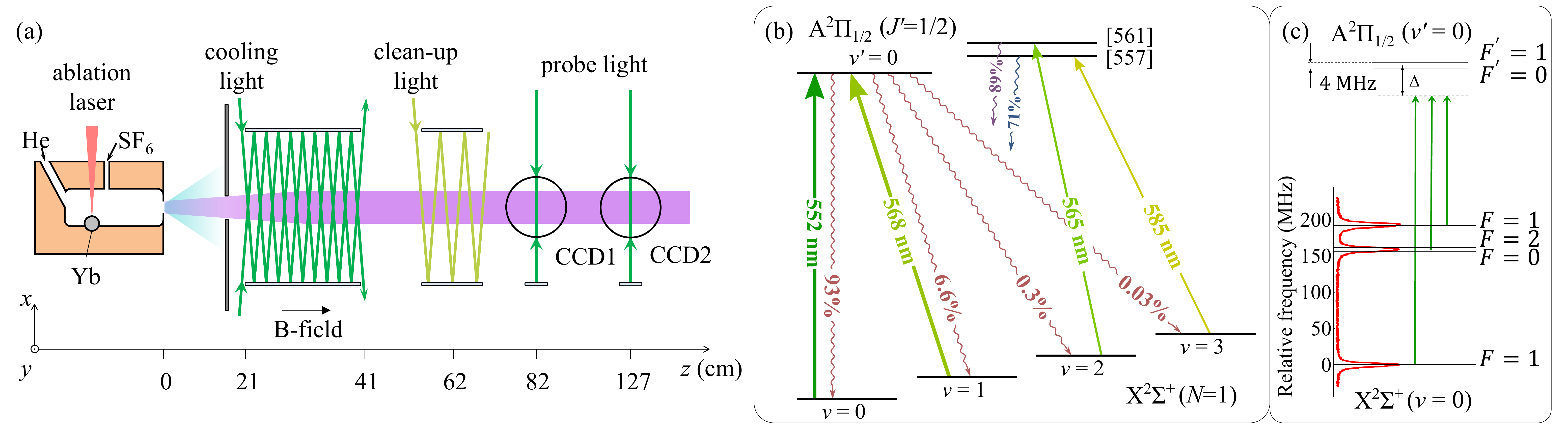}
		\caption{\label{figure1} (Color online) (a) Schematic of the experiment. (b) Relevant energy levels and branching ratios. Solid arrows: transitions used for laser cooling and repumping, along with their wavelengths. Wavy arrows: Spontaneous decays with their branching ratios. The states labelled [561] and [557] are mixtures of the $A^{2}\Pi_{1/2}(v'=1)$ state and a perturbing state with $\Omega=1/2$, sometimes called  [18.6]0.5~\cite{Sauer1999}. For these states, the branching ratios of the dominant decays to $v=1$ are shown. (c) Hyperfine structure of the main cooling transition, $X^{2}\Sigma^{+}(v=0, N=1)$--$A^{2}\Pi_{1/2}(v'=0, J'=1/2)$. The frequency components of the 552~nm laser are indicated by the spectrum in red. For red detuning (as shown here)  $\Delta$ is negative.}
	\end{figure*}

Figure~\ref{figure1}(a) illustrates the experiment. At $z=0$, pulses of YbF are emitted from a cryogenic buffer gas source of the same design as Ref.~\cite{Truppe2017c}. The pulses have a duration of 250~$\mu$s, a mean forward speed of $v_{z} \simeq 160$~m/s, and a $^{174}$YbF flux of $5{\times}10^9$ molecules per steradian per pulse in the first rotationally-excited state. After a 4~cm diameter aperture at $z=20$~cm, the molecules pass through the 20-cm-long laser cooling region, the 5.5-cm-long clean-up region, and the detectors placed at distances $l_1=41$~cm and $l_2=86$~cm from the end of the cooling region. A magnetic field $B$, applied in the $z$-direction, is uniform to within 0.1~G throughout the cooling region. Additional shim fields cancel the background magnetic field.

Figure~\ref{figure1}(b) shows the relevant energy levels of $^{174}$YbF and the branching ratios between them~\cite{Smallman2014, Zhuang2011}. The main laser-cooling transition is the rotationally-closed $X^{2}\Sigma^{+}(v=0, N=1)$--$A^{2}\Pi_{1/2}(v'=0,J'=1/2)$ transition at 552~nm, with a linewidth of $\Gamma=2{\pi}\times 5.7$~MHz, and single-photon recoil velocity of 3.7~mm/s. Additional lasers repump molecules that decay to the $v=1,2,3$ vibrational states of $X^{2}\Sigma^{+}$, implementing the scheme proposed in Ref.~\cite{Smallman2014}. Each ground (excited) state has 4 (2) hyperfine components, all with differing intervals~\cite{Lim2017}. Those of the main cooling transition are shown in Fig.~\ref{figure1}(c). Acousto-optic and electro-optic modulators add to each laser beam the radio-frequency sidebands required to excite all ground hyperfine levels. These beams are coupled into a single-mode polarisation-maintaining optical fibre that delivers the light to the molecules. The fibre output, with typical powers ($P_{v}$) of 50, 170, 18 and 6~mW in the four wavelengths addressing $v=0,1,2,3$, is collimated to give a Gaussian intensity distribution having 4.4~mm $1/e^{2}$ diameter. The light is split into two beams of equal intensity, which enter the cooling region from opposite sides and cross back and forth 38 times in the $xz$-plane between parallel mirrors. The beams are linearly polarized, one at an angle of $\pi/4$ to the $y$-axis and the other at $\pi/4 + \phi$. Each laser, with sidebands added, is separately tuned to produce the maximum fluorescence in the laser cooling region. The main cooling laser is then detuned by the angular frequency $\Delta$, with sub-Doppler cooling expected for $\Delta > 0$.

The light in the cleanup region has all the repump frequencies and none of the 552~nm cooling frequencies, so any population remaining in the $X^{2}\Sigma^{+}(v=1,2,3)$ states is driven to $X^{2}\Sigma^{+}(v=0, N=1)$. The spatial distribution of molecules in this state is then measured by recording laser-induced fluorescence on one of two CCD cameras. The 552~nm probe light used in these detectors is the same as the cooling light, but is independently tuned to $\Delta=0$. This light crosses the molecular beam at right angles and is retro-reflected.

\begin{figure*}[tb]
 \centering
 \includegraphics[width=1\textwidth]{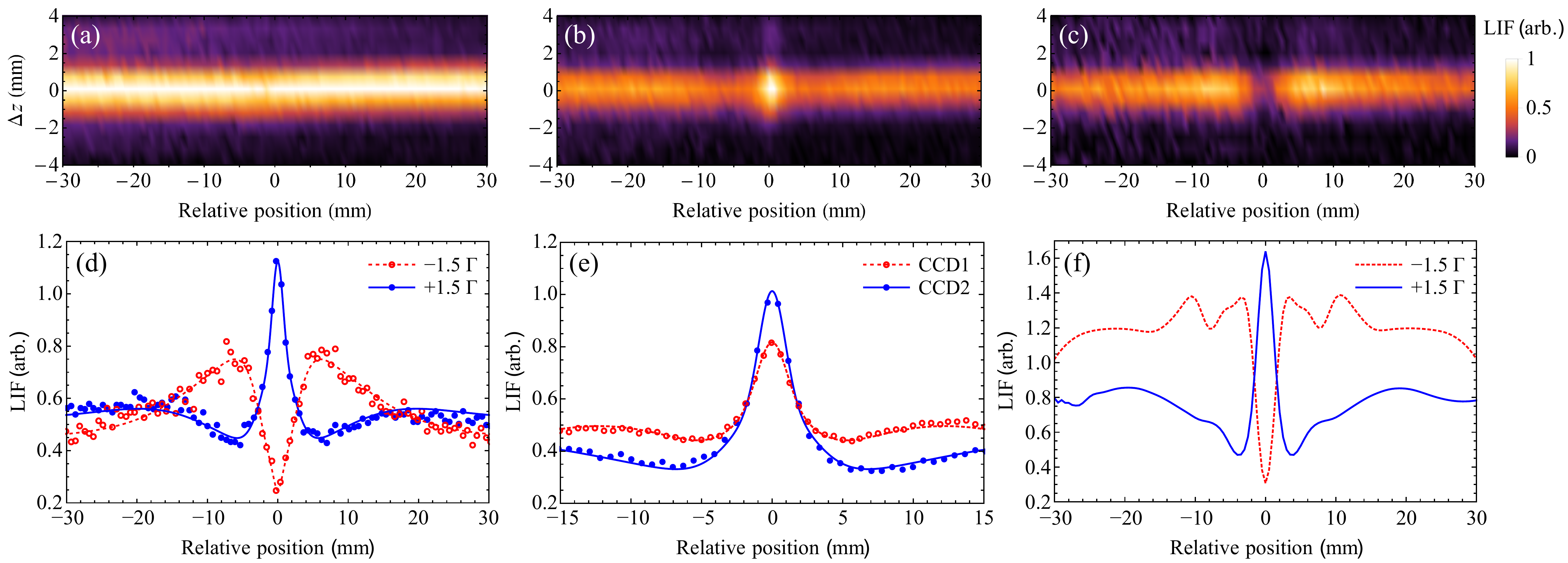}
\caption{\label{figure2} (a,b,c) Fluorescence images at CCD2 with $B=0.8$~G, $\phi=0$ and: (a) no light in the cooling region, (b) $\Delta=+1.5 \Gamma$, (c) $\Delta=-1.5 \Gamma$. (d) Normalized density distributions along $x$ obtained by integrating the images over the $z$-direction and then dividing (b) \& (c) by (a). (e) Normalized density distributions at CCD1 (red open circles) and CCD2 (blue solid circles), for $\Delta=2 \Gamma$, $B=1.2$~G, and $\phi=0$. Here, $P_{0}=90$~mW. Lines are fits to the four-Gaussian model discussed in the text. (f) Simulation results for the parameters corresponding to (d). These simulated distributions conserve the number of molecules once integrated over a wider range of positions than shown.}
\end{figure*}

Figure~\ref{figure2}(a-c) shows images obtained from CCD2 when $B=0.8$~G and $\phi=0$. In (a) the cooling light is absent and we observe a uniform fluorescence intensity across the field of view in the $x$-direction, reflecting the uniform density of molecules and uniform detection efficiency over this region. The variation of fluorescence along $z$ reflects the intensity distribution of the probe laser. In (b) the cooling is applied with $\Delta=+1.5\Gamma$ and we see a bright spot at the centre while in (c), $\Delta=-1.5 \Gamma$, and we observe a hole at the centre. To analyse these images we first integrate them along $z$, over the 8~mm range shown, and then divide the distribution with cooling applied (b, c) by the one with no cooling (a), giving the normalised fluorescence distributions shown in figure \ref{figure2}(d). The narrow peak obtained when $\Delta=+1.5\Gamma$ is due to magnetically-assisted sub-Doppler cooling (see below) which cools slow-moving molecules to low temperature, producing a beam with a highly collimated centre. Further out, there is a dip where molecules that form the peak would otherwise have been. When $\Delta = -1.5\Gamma$, there is a dip at the centre with broad wings on either side because the same mechanism now drives slow molecules to higher speeds. Doppler cooling also contributes to these broad wings by reducing the velocity of molecules at higher velocities. The distance between the minima (at $\pm x_{\rm min}$) for $\Delta = +1.5\Gamma$, or between the maxima for $\Delta=-1.5\Gamma$, increases from CCD1 to CCD2. From this change, we infer an approximate capture velocity for the sub-Doppler cooling process of $v_{c} \approx 0.9$~m/s.  Fewer molecules are detected in total when the cooling light is applied. This depletion reaches its largest value, 55\%, when $\Delta=0$. We have investigated and ruled out several possible causes for this loss, including deflection or heating of the molecular beam in the $y$-direction, incomplete optical pumping in the clean-up region, or decay to other rotational levels. We are currently investigating unexpected losses to higher-lying vibrational states.

We fit the density distributions to a sum of four Gaussians, $G_{i}$, all having a common centre, but with differing amplitudes and widths. For $\Delta >0$, $G_{1}$ represents the narrow central peak, $G_{2}$ the broad dip, and $G_{3}$ the even broader curvature of the baseline. $G_{4}$, always of low amplitude, helps to reproduce the shoulders of the narrow peak. Fits to the data in Fig.~\ref{figure2}(d) are shown by the lines. We define the peak height and peak width as the amplitude and width of $G_{1}$, and use these parameters to quantify the effectiveness of the cooling.

Figure~\ref{figure2}(e) shows the normalised density distributions at the two CCDs when $\Delta=2 \Gamma$, $B=1.2$~G, $\phi=0$ and $P_{0}=90$~mW. The peak is higher at CCD2 because the density of the uncooled beam, which provides the normalisation, decreases faster than that of the cooled beam. Our simulations (see below) show that molecules in the central peak have a Boltzmann velocity distribution with no correlation between position and velocity. In this case, the temperature is given by
\begin{equation}
\label{temperature}
T=\frac{m{v_z}^2}{k_{\rm B}}\frac{{w_2}^2-{w_1}^2}{{l_2}^2-{l_1}^2},
\label{Eq:Temperature}
\end{equation}
where $w_{1,2}$ are the rms widths of the peaks measured at the two detectors. Fitting the four-Gaussian model to the data in Fig.~\ref{figure2}(e) gives $w_{1} = 1.009 \pm 0.045$~mm and $w_{2} = 0.905 \pm 0.021$~mm, where the errors are the statistical uncertainties from the fit. These uncertainties translate into a temperature resolution of 100~$\mu$K. The systematic error due to uncertainty in the imaging magnification and various misalignments of the cameras are below 0.6\%, so contribute negligibly to the uncertainty. Although $w_{2}<w_{1}$, implying a negative temperature, the widths differ by only $2\sigma$. We conclude that the temperature is below the resolution of the measurement, giving us an upper temperature limit of $T_{\rm upper} = 100 \mu$K. Given the measured widths, the probability of the temperature being higher than this is only 0.13\%. Fits to unnormalised data also give $T$ consistent with zero and an upper limit reduced to 80~$\mu$K. Fitting single Gaussians to the data lying within $x<x_{\rm cut}$ gives $T$ consistent with zero, with an uncertainty below 100~$\mu$K, for all choices of $x_{\rm cut} < x_{\rm min}$. We have modelled the possibility that the central peak contains two distributions, one at temperature $T_{\rm hot}$, and the other at a much lower temperature, $T_{\rm cold}$. This can lead to $w_2 < w_1$ because the hot, rapidly expanding component broadens the peak at the first detector more than at the second. The model can reproduce the observed widths provided $T_{\rm hot}>0.5$~mK and $T_{\rm cold} < 35~\mu$K. Though this method lowers the temperature limit, it depends on the model being the correct one, so we prefer to use $T_{\rm upper}$ as a more conservative upper limit. Our estimate of $T_{\rm upper}$ is below the Doppler temperature, which for our parameters is $T_{\rm D}\approx450~\mu$K. It is also below the minimum Doppler temperature of $T_{\rm D, min}=\hbar\Gamma/(2 k_{\rm B})=137~\mu$K.

Figure \ref{figure2}(f) shows the results of simulating these experiments. The cooling force and momentum diffusion coefficient are calculated as a function of velocity by solving the optical Bloch equations, following the approach described in Ref.~\cite{Devlin2016} but extended to account for the hyperfine structure of the ground and excited states and the three frequency components of the 552~nm light. Using these results, the distribution of molecules is calculated by solving the Fokker-Plank equation \cite{Molmer1994}. For $\Delta = 1.5\Gamma$, the simulations reproduce the cooling data well, showing a central peak of width 1.17 mm, similar to the measured width of $0.87 \pm 0.04$~mm. The simulations predict a capture velocity for sub-Doppler cooling of $v_{c} = 0.6$~m/s, close to the value estimated above. For $\Delta = -1.5\Gamma$, the simulations reproduce the central dip seen in the experiment, but show additional structure either side of the peak which is not seen experimentally. For the parameters corresponding to Fig.~\ref{figure2}(e), the simulations predict a temperature of $1 \mu$K, only six times higher than the recoil temperature and far below the temperature resolution of the experiment. Artificially reducing the cooling force or increasing the momentum diffusion constant by a factor of 10 increases the predicted temperature to $10 \mu$K.

\begin{figure}[tb]
	\centering
	\includegraphics[width=0.46\textwidth]{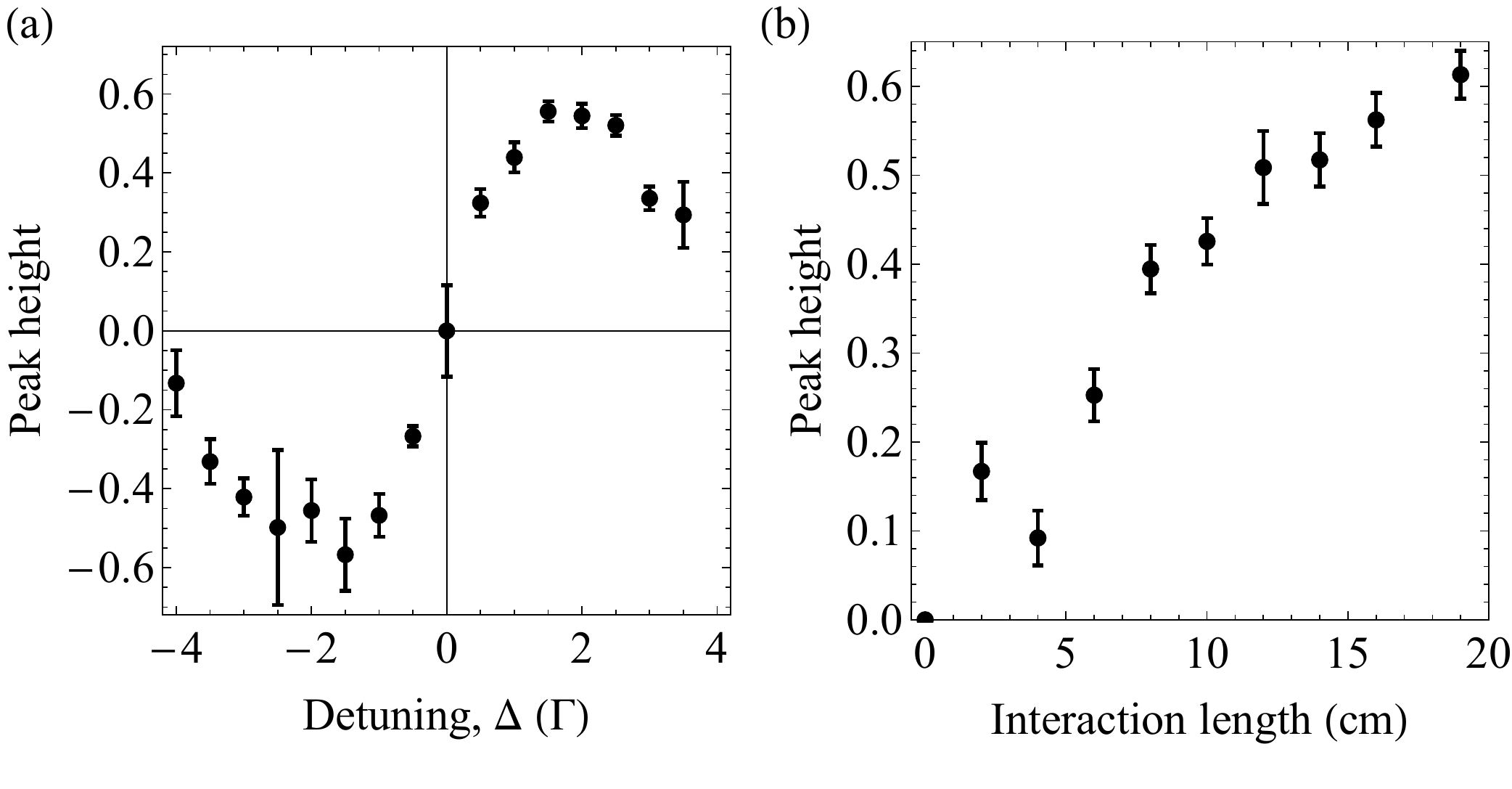}
	\caption{\label{figure3}(a) Peak height versus $\Delta$, when $B=0.8$~G, $\phi=0$ and $L=20$~cm. (b) Peak height versus the length of the cooling region, $L$, when $B=0.8$~G, $\phi=0$ and $\Delta$=+1.5 $\Gamma$. Error bars are the standard errors obtained from the fits.}
\end{figure}

With $\Delta>0$, the width of the narrow peak varies little for the range of parameters explored, and the temperature is too low to measure. However, the peak height varies strongly with the parameters and is a good measure of the number of utracold molecules. Figure~\ref{figure3}(a) plots the peak height versus $\Delta$ showing the dispersive shape characteristic of laser cooling. The height is symmetric about $\Delta = 0$ and is largest when $\Delta \simeq +2 \Gamma$. Figure~\ref{figure3}(b) shows that the peak height increases with the length of the cooling region, $L$, but starts to level off once $L \approx 20$~cm, the maximum value explored here.

\begin{figure}[!t]
	\centering
	\includegraphics[width=0.46\textwidth]{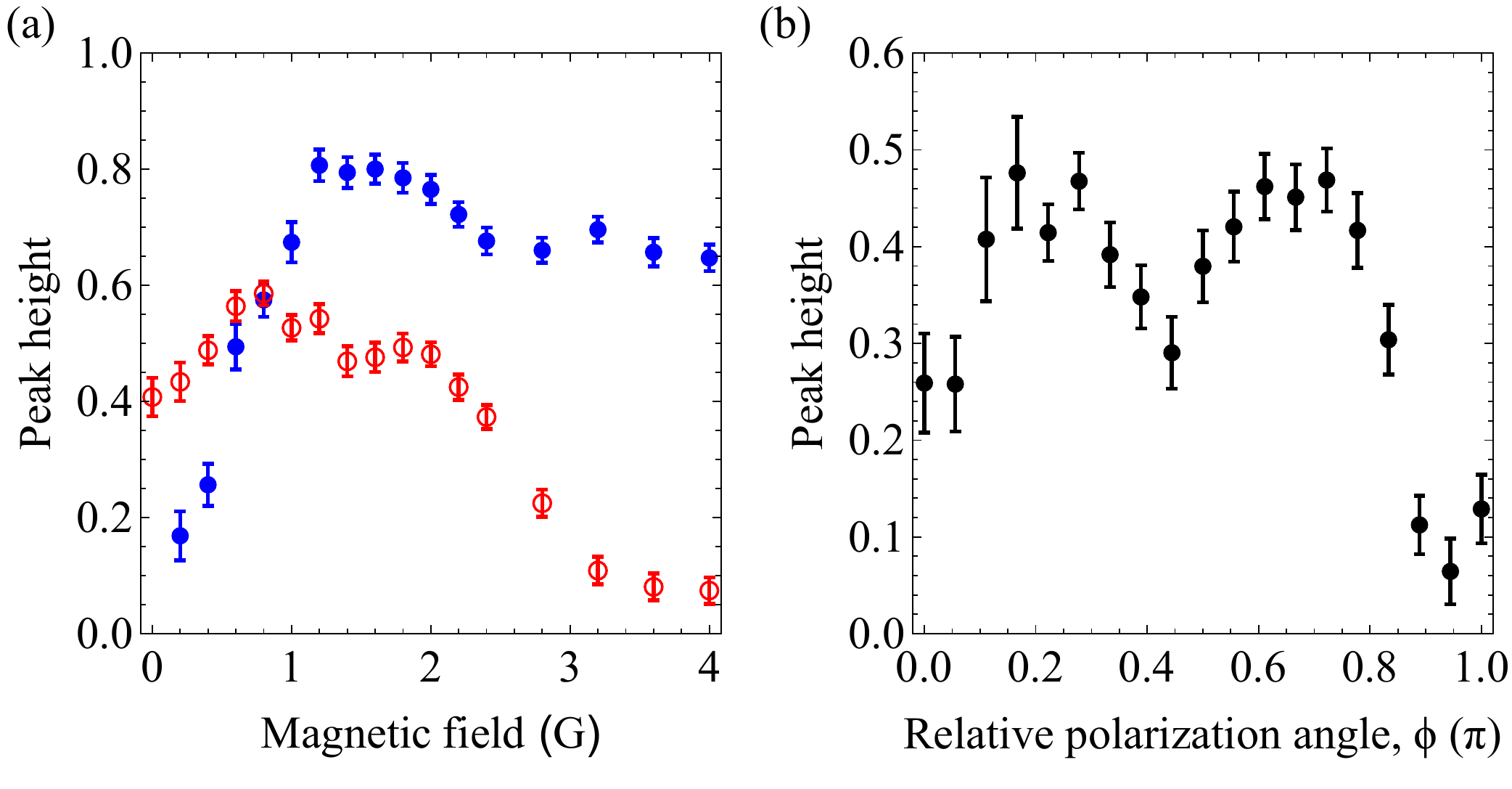}
	\caption{\label{figure4} (a) Peak height versus $B$ when $\Delta=2 \Gamma$ for two cases: $\phi=0$ (filled blue points) and $\phi=\pi/2$ (open red points). (b) Peak height vs $\phi$ when $\Delta=+2\Gamma$ and $B=0$. Error bars are the standard errors obtained from the fits.}
\end{figure}

The sub-Doppler cooling demonstrated here relies on the presence of dark states, and the mechanism depends on the polarization configuration, which we control through $\phi$~\cite{Devlin2016}. When $\phi = 0$, the polarization is uniform but there are standing waves of intensity. A molecule in a bright state, moving towards high intensity, climbs the potential hill arising from the ac Stark shift and is optically pumped into a dark state near the top of the hill. As it moves on towards a region of low intensity, the applied magnetic field rotates the dark state back into a bright state. Thus, molecules spend most of their time climbing potential hills. This mechanism, known as magnetically-induced laser cooling~\cite{Ungar1989, Sheehy1990, Emile1993, Gupta1994}, has been used for transverse cooling of SrF and SrOH beams~\cite{Shuman2010, Kozyryev2017}. The filled points in Fig.~\ref{figure4}(a) show the peak height versus $B$ for this $\phi = 0$ case. We see that the cooling is ineffective if $B$ is too small, and that the peak height increases with $B$ up to $B \approx 1.2$~G. Cooling should be optimized when the Larmor precession time is about the same as the time taken for a molecule to move from a node to an antinode of the standing wave. We saw above that the cooling is effective for speeds up to $v_{c} \approx 0.9$~m/s. Taking $v_{c}/2$ as a typical speed, we expect an optimum $B$ of $B_c \approx 2\hbar v_{c}/(g \mu_B \lambda)$. Averaging the $g$-factors of the various ground-state hyperfine components, weighted by their degeneracy, gives $g=1/3$ and $B_{c} \approx 1.1$~G, matching the optimum found experimentally. For higher $B$, the peak height is surprisingly insensitive to $B$, perhaps because of the wide range of $g$-factors and molecule speeds involved. We find the cooling to be effective up to $B \approx 15$~G. 

When $\phi = \pi/2$, the intensity is uniform but the polarization is not, and the sub-Doppler mechanism involves non-adiabatic transitions between dark and bright states induced by motion through this changing polarization~\cite{Weidemuller1994, Fernandes2012}. This mechanism does not require a magnetic field. The open points in Fig.~\ref{figure4}(a) show the peak height versus $B$ when $\phi = \pi/2$. We see that the cooling is effective at $B=0$, as expected since this mechanism does not require a magnetic field, unlike the $\phi = 0$ case. On the contrary, the data show that magnetic fields above 2~G are detrimental to this cooling mechanism. 

Figure~\ref{figure4}(b) shows how the peak height depends on $\phi$ when $B=0$. The data are roughly symmetric around $\phi=\pi/2$ as we would expect, and the peak height is largest near $\phi=\pi/4$ and $3\pi/4$, and smallest near $\phi=0$ and $\pi/2$. Simulations for this polarization configuration~\cite{Devlin2016}, called lin-$\phi$-lin, show that the sub-Doppler force is maximised for $\phi$ between $\pi$/8 and 3$\pi$/16 for the $1\rightarrow1$ system and between 3$\pi$/16 and $\pi$/4 for the $2 \rightarrow 1$ system. Our measurements are consistent with those results. 

In summary, we have cooled YbF molecules to sub-Doppler temperatures by realizing the laser cooling scheme proposed in \cite{Smallman2014}, and have explored how the cooling efficiency depends on the main parameters. This is a key step towards using ultracold molecules for an eEDM measurement~\cite{Tarbutt2013}, and other tests of fundamental physics. Our temperature limit of $T < 100~\mu$K extends the feasible coherence time to $\tau > 150$~ms. To make use of this with the current $v_z$ would require a 24~m-long experiment. A slower beam could be obtained by radiation pressure slowing~\cite{Barry2012,Truppe2017,Hemmerling2016}, or by a potentially more efficient approach such as Zeeman-Sisyphus slowing~\cite{Fitch2016} or bichromatic force slowing~\cite{Kozyryev2017c}. For the data shown in Fig.~\ref{figure2}(e), there are about 1.3$\times 10^4$ molecules in the ultracold part of the distribution. We are currently extending the method into 2D, which should yield far more molecules, especially since the capture velocity for sub-Doppler cooling is larger in 2D~\cite{Devlin2016}. The combination of a Doppler cooling period followed by sub-Doppler cooling would increase the capture velocity further while providing the same low final temperature. 

This work was supported by the STFC and has received funding from the European Research Council under the European Union's Seventh Framework Programme (FP7/2007-2013)/ERC grant agreement 320789.

\bibliography{references}
\end{document}